\documentclass{emulateapj}
\usepackage{graphicx,natbib}
\usepackage{bm}
\usepackage{epsfig}
\usepackage{color}

\def\hideprivate{
   \global\long\def\private##1{
   }
}

\hideprivate    
\shorttitle{Precision cosmology from gravitational waves}
\shortauthors{Nissanke et al.}

\begin{document}

\title{Determining the Hubble constant from gravitational wave observations of merging compact binaries}

\author{Samaya Nissanke\altaffilmark{1}, Daniel
  E. Holz\altaffilmark{2}, Neal Dalal\altaffilmark{3}, Scott
  A. Hughes\altaffilmark{4,5,6}, \\ Jonathan L. Sievers\altaffilmark{7,8}, Christopher M. Hirata\altaffilmark{1}}

\altaffiltext{1}{Caltech, Theoretical Astrophysics, California
  Institute of Technology, Pasadena, California 91125, USA}

\altaffiltext{2}{Enrico Fermi Institute, Department of Physics, and
  Kavli Institute for Cosmological Physics, University of Chicago,
  Chicago, IL 60637, USA} 

\altaffiltext{3}{Department of Astronomy, University of Illinois, 1002
  W. Green St., Urbana, IL 61801, USA}

\altaffiltext{4}{Department of Physics and MIT Kavli Institute, MIT, 77
 Massachusetts Ave., Cambridge, MA 02139, USA}

\altaffiltext{5}{Canadian Institute for Theoretical Astrohysics,
 University of Toronto, 60 St.\ George St., Toronto, ON M5S 3H8,
 Canada}

\altaffiltext{6}{Perimeter Institute for Theoretical Physics, Waterloo, ON
 N2L 2Y5, Canada}

\altaffiltext{7}{Jadwin Hall, Department of Physics, Princeton
  University, New Jersey, USA}

\altaffiltext{8}{Astrophysics and Cosmology Research Unit, University of Kwazulu-Natal, Westville, Durban, 4000, South Africa}

\begin{abstract}
Recent observations have accumulated compelling evidence that some short
gamma-ray bursts (SGRBs) are associated with the mergers of neutron star (NS)
binaries. This would indicate that the SGRB event is associated with a gravitational-wave (GW) signal corresponding to the
final inspiral of the compact binary. In addition, the radioactive decay of
elements produced in NS binary mergers
may result in transients visible
in the optical and infrared with peak luminosities on hours-days timescales. Simultaneous observations of the
inspiral GWs and signatures in the electromagnetic band may allow us to directly and
independently determine both the luminosity distance and redshift to a binary. These standard
sirens (the GW analog of standard candles) have the potential to provide an accurate measurement of the
low-redshift Hubble flow. In addition, these systems are absolutely
calibrated by general relativity, and therefore do not experience the
same set of astrophysical systematics found in traditional standard candles, nor
do the measurements rely on a distance ladder. We show that 15
observable GW and EM events should allow the Hubble constant to be
measured with 5\% precision using a network of detectors that includes
advanced LIGO and Virgo.  Measuring 30 beamed GW-SGRB events could
constrain $H_0$ to better than 1\%. When comparing to standard
Gaussian likelihood analysis, we find that each event's non-Gaussian posterior in $H_0$ helps reduce the overall measurement errors in $H_0$ for an ensemble of NS binary
mergers.
\end{abstract}

\keywords{cosmology: distance scale---cosmology: theory---gamma rays:
bursts---gravitational waves}


\section{Introduction}


Gravitational wave (GW) standard sirens are the GW analogs to traditional
standard candles. They exemplify multi-messenger astronomy (see
\citealt{bloometal09}, \citealt{phinney09}, \citealt{kk:2009}), where the
use of both electromagnetic (EM) and GW measurements results in astrophysical
insights inaccessible to either method alone. As originally discussed by
\cite{schutz86}, inspiralling compact neutron star (NS) or black hole (BH)
binaries are excellent standard sirens, in that their GW measurements
could determine the sources' absolute distances.  The only calibration
(so to speak) in this measurement is the assumption that general
relativity describes the binary waveform. When used in conjunction
with a redshift measurement inferred by an independently observed EM counterpart,
standard siren observations allow us to study the luminosity distance-redshift
relationship. Consequently we may map out the Universe's expansion history, and
thereby constrain cosmological parameters such as the Hubble constant
$H_0$, the dark energy equation-of-state parameter $w$, and the average densities of
matter $\Omega_m$ and dark energy $\Omega_{\lambda}$.

In this work we are interested in the constraints on the Hubble constant that
result from observations of GW-EM standard sirens at low redshift ($z < 0.3$),
as would be expected from the coming generation of ground-based GW observatories.
This will complement existing $H_0$ measurements at the few percent
level, which have been determined using a combination of methods; see
\cite{Suyu:2012} for a brief review and references therein. All
methods aim to i) provide independent $H_0$ measurements, compared to derived constraints in the case
of the Cosmic Microwave Background (CMB) and Baryon Acoustic
Oscillations (BAO) measurements (see, for example, discussion of cosmological parameter constraints by the \citealt{Planck:2013}), and ii) reach a $\sim$ 1\%
precision in random errors in the near future. For instance, the
Planck satellite recently reported $H_0$ of
67.4 $\pm$ 1.4 km s$^{-1}$ Mpc$^{-1}$,  a low value compared to cosmic
distance ladder results of 73.8 $\pm$ 2.4 km s$^{-1}$ Mpc$^{-1}$ (HST
Cepheid--SNIa; \citealt{Riess:2011}a, \citealt{Riesserr:2011}b) and 74.3 $\pm$ 1.5 [statistical] $\pm$ 2.1 [systematic] km s$^{-1}$ Mpc$^{-1}$
(Spitzer CHP; \citealt{Freedman:2012}). Although the discrepancy between the CMB and
cosmic distance ladder measures lies only at a $\sim$ 2.5 $\sigma$
discrepancy, such differences suggest the necessity of additional independent measures. An independent $\sim 1$\%
measurement in $H_0$ is especially desirable when improving figure-of-merit
constraints on the dark energy equation of
state~\citep{2005ASPC..339..215H,Weinberg2012}. The control of systematics is crucial in having a useful measurement
  of $w$, especially one that might involve the falsification of a cosmological
  constant as the origin of the dark energy. Standard siren approaches offer a
  fundamentally different set of systematics, and therefore provide a valuable
  counterpart to all other methods.

For the next generation of advanced ground based GW observatories, comprising 
LIGO\footnote{http://www.ligo.caltech.edu/advLIGO/},
Virgo\footnote{http://www/ego-gw.it/public/virgo/virgo.aspx}, and other
GW interferometers, inspiralling and
merging NS-NS and NS-BH binaries are expected to be the most numerous and
characterizable events. For an advanced
LIGO--Virgo network, predicted event rates for NS--NS binary mergers range from
$0.4$ to $400$ year$^{-1}$, and such systems could be detectable to distances of several
hundred Mpc \citep{Abadieetal:2010,aasi:2013}. Similar rate estimates
apply to NS-BH binaries.  These numbers are, however, far more
uncertain (by several orders of magnitude) since they rely on
population synthesis alone \citep{Abadieetal:2010}. NS-BH binaries
are detectable to much larger distances ($>1$ Gpc, depending on the
mass of the BH) thanks to their larger masses (which leads to a higher
amplitude GW). Short-hard gamma-ray bursts (SGRBs) are believed to originate in the
mergers of NS binaries (e.g., \citealt{Fong:2013} and references therein), and could therefore provide GW
standard sirens with EM counterparts (\citealt{dalaletal}, hereafter
D06). In addition, another potentially important standard siren is afforded
by ``macronovae'' or ``kilonovae'' (e.g., \citealt{Li:1998,Kulkarni:2005,Metzger:2010,Roberts:2011,Barnes:2013,Kasen:2013}), which are weak supernova-like
transients with isotropic emission fueled by radioactive powered ejecta resulting from the mergers of NS--NS and
NS--BH binaries.

The D06 analysis considered measurements using an advanced
four-detector network that included an Australian detector (whose
construction was until fairly recently under consideration).  It also
used a parameter estimation formalism based on a Gaussian
approximation to the posterior likelihood function.  This approximation is
strictly correct only for ``high'' signal-to-noise ratio (SNR),
although it is not clear what ``high'' means in this context (\citealt{cf94,vallisneri08}).  This is troublesome for
ground-based GW detectors where most measurements are expected to be
low SNR (near threshold).  As a consequence, D06 {\it underestimates}
the distance errors that we expect.

In \citet{NHHDS10} (hereafter N10) we revisited the D06 analysis by following a
Bayesian approach for estimating parameter errors instead of
a Fisher matrix analysis (\citealt{finn92}). We implemented a
Metropolis-Hastings Markov Chain Monte--Carlo (MCMC) method to explore the
posterior distribution of GW model parameters, in particular for deriving measurement
errors in the luminosity distance. This was done for both NS--NS and
NS--10 $M_{\odot}$ BH binaries, using a careful selection procedure to decide which
binaries to include in the analysis. N10 also used a more accurate inspiral
waveform than was used in D06, and considered various Northern and Southern Hemisphere ground-based detector
networks. Specifically, in N10 we examined measurement accuracies at different network
combinations of LIGO Hanford, LIGO Livingston, Virgo, KAGRA (formally
known as the ``Large-scale
Cryogenic Gravitational-wave
Telescope'')\footnote{http://gw.icrr.u-tokyo.ac.jp:8888/lcgt/}, and
AIGO/LIGO Australia \citep{LIGOAustralia}. We also considered how
beaming of the SGRB, derived from observations using present or future
$\gamma$-ray observatories, reduces measurement errors in the luminosity
distance. In N10 we found that the distance to an individual NS-NS binary is measured to within a fractional error of
roughly $20$--$60$\%, with $20$--$30$\% being representative for the
majority of events in our detected distribution. If the orientation of
the orbital plane of the NS-NS binary is assumed to be face-on (as might be expected
for beamed SGRBs), we found that distance measurement errors
improve by approximately a factor of two. If we instead assume that
the EM counterparts are NS--10M$_{\odot}$ BH mergers, we found
that the distribution of fractional distance errors ranges from $15$--$50$\%, with most
events clustered near $15$--$25$\%. Assuming that the EM counterpart
is a beamed SGRB reduces the measurement errors by a factor of two.

In what follows we update and refine the analysis of D06 and N10. This paper focuses on the implications such measurements will have when constraining
cosmological parameters such as $H_0$.  Subtle and important
differences exist between our current analysis and that used in
N10. Primarily, we are interested here in $H_0$ constraints for the
ensemble of GW-EM events, and not in individual distance measures for
GW-EM event as in N10. In addition to collimated SGRBs, we
also consider more speculative transients, such as macronovae or kilonovae,
associated with NS binary mergers that emit isotropically in the optical or near
infrared. For consistency with N10, we assume that the BHs in our
NS-BH populations have masses of
10M$_{\odot}$. We note, however, that recent numerical relativity
simulations suggest that tidal disruption, and hence EM signatures,
may only occur for NS-BH mergers with much smaller BH masses ($\sim$ 5
M$_{\odot}$); see e.g., \cite{Taniguchi:2007}, \cite{Shibata:2008},
\cite{Shibata:2009}, \cite{Kyutoku:2011}, \cite{Foucart:2011}, \cite{Foucart:2012}. In
addition, we examine measurement
accuracies for networks including LIGO
India, an advanced interferometer whose construction is currently
under consideration \citep{LIGOIndia}. Finally, in contrast to traditional
standard candles such as Cepheid variables and Type Ia supernovae, we wish to
emphasize that GW standard sirens are independent of the cosmological distance
ladder. Compared with other recent standard siren studies using advanced GW
interferometers (see e.g., \citealt{DelPozzoPRD:2012,Taylor:2012,Messenger:2012}), we consider the
case where an EM observation of the NS binary inspiral is seen in
conjunction with a GW measurement. 

In the next section we summarize the principles
underlying GW standard sirens. We then outline how we select our sample of
binaries, and discuss the Bayesian method employed when estimating the luminosity
distance for each source. We conclude by discussing future constraints on
 $H_0$, the results of which are critically dependent on the assumed
source population's characteristics and the specific advanced detector network.

\section{Standard Siren Binaries}

Many key observational methods employed in mapping out the expansion history
of the universe rely on the luminosity distance-redshift relation:
\begin{eqnarray}
D_L(z)&=&\frac{c \, (1+z)}{H_0\sqrt{\Omega_K}} \sinh\left[\sqrt{\Omega_K}
\int_0^{z} \, \frac{H_0}{H(z')} \, dz'\right],
\label{eqn:DLz}
\end{eqnarray}
where the luminosity distance $D_L(z)$ is given as a redshift
integral of the Hubble parameter $H(z)$, and the Hubble constant
$H_0$.  For $z\gtrsim 1$, the evolution of $H(z)$ and $D_L(z)$ depends
on cosmological parameters like $\Omega_m$ and $w$, through the
Friedmann equations.  However, for low redshifts $z\ll 1$, the
distance-redshift relation is well described by $D_L(z) \approx
c\,z/H_0$, independent of other cosmological parameters. This is why
measurements of the distances to local sources, like Cepheids or
GW standard sirens, can constrain the value of the Hubble constant.

The inspiral signal of the GWs, modeled accurately using the
post-Newtonian (PN) approximation in general relativity, encodes geometrical and physical parameters of the source (see e.g.,
\citealt{blanchet06}). The source parameters include: the binary's
luminosity distance $D_L$, its position on the sky $\mathbf{n}$, its
redshifted chirp mass ${\cal M}_z = (1+z)
\, m_1^{3/5} m_2^{3/5} / (m_1+m_2)^{1/5}$ where $m_1$ and $m_2$ denote
the mass of each compact object in the binary, its redshifted reduced
mass $\mu_z = (1+z) \, m_1 m_2/ (m_1+m_2)$, its orientation on the sky given by its
inclination angle $\iota$, where $\cos\iota = \mathbf{L} \cdot \mathbf{n}
/ |\mathbf{L}|$ and $\mathbf{L}$ is the binary's
orbital angular momentum, and $t_c$ and $\Phi_c$, the time and GW phase at
merger. A single detector $a$ measures a linear combination of the two
GW polarizations: 
\begin{equation}
h_{a,{\rm meas}}(\bf{\theta}) =  F_+(\theta,\phi,\psi) \, h_+ +
F_\times(\theta,\phi,\psi) \, h_\times\;.
\end{equation}
\noindent The colatitude $\theta$ and longitude $\phi$ describe the binary's
position on the sky ${\bf n}$. The polarization angle
$\psi$ sets the inclination of the components of the unit vector $\mathbf{\hat L}$ orthogonal to the unit
vector $\mathbf{\hat n}$. The components of the vector
$\mathbf{\theta}$ are all the various parameters (masses, angles,
distance, etc.) upon which this measured waveform depends. For the two GW polarizations $h_+$ and $h_{\times}$, we use the
non-spinning restricted 3.5PN waveform in the frequency domain
(indicated by the $\tilde{h}$ notation):
\begin{eqnarray}
\tilde{h}_+(f) &=& \sqrt{\frac{5}{96}}\frac{\pi^{-2/3} {\mathcal
M}_z^{5/6}}{D_L} [1+(\mathbf{\hat{L}}\cdot \mathbf{\hat{n}})^2]   f^{-7/6} e^{i\Psi(f)} \, ,
\label{eq:freqdomainhp}
\\
\tilde{h}_\times(f) &=& \sqrt{\frac{5}{96}}\frac{\pi^{-2/3} {\mathcal
M}_z^{5/6}}{D_L}(\mathbf{\hat{L}}\cdot \mathbf{\hat{n}})  f^{-7/6} e^{i\Psi(f) - i\pi/2} \, ,
\label{eq:freqdomainhc}
\end{eqnarray}
which relies on the ``stationary phase'' approximation \citep{fc93}, where
the GW frequency $f$ varies slowly over a single wave period. The GW phase in the
frequency domain $\Psi$ is computed to
3.5 PN order, where $\Psi (f)$ is given by:
\begin{eqnarray}
\Psi(f) & = & 2\pi f t_c - \Phi_c - \frac{\pi}{4} + \frac{3}{128}(\pi {\mathcal M}_z f)^{-5/3}
\times \nonumber \\ & & \left[ 1 + \frac{20}{9} \left( \frac{743}{336} +
\frac{11}{4}\eta \right) (\pi M_z f)^{2/3} \right. \nonumber \\ & & \quad -16\pi( \pi M_z f) + 10 \left( \frac{3058673}{1016064} +
\frac{5429}{1008}\eta \right.  \nonumber \\ & &  \left. +
\frac{617}{144}\eta^2 \right)(\pi M_z f)^{4/3} \, + \, \pi \left(
\frac{38645}{756} - \frac{65}{9} \eta \right) \nonumber \\ & & 
\times \left[ 1 + (\pi M_z f)^{5/3} \ln \left(\frac{f}{f_0}\right) \right]
\nonumber \\ & & + \left[ \frac{11 583 231 236 531}{4 694 215 680} -
  \frac{640}{3} \pi^2 - \frac{6 848}{21} \gamma \right] (\pi M_z f)^{2} \nonumber \\ & & \left[ \left( - \frac{15 335 597 827}{3 048
      192} + \frac{2 255}{12} \pi^2 - \frac{47 324}{63} - \frac{7
      948}{9} \right) \eta \right. \nonumber \\ & & \left. + \,  \frac{76 055}{1 728} \eta^2 - \frac{127 825}{1 296} \eta^3 \right]
  (\pi M_z f)^{2} \nonumber \\ & & + \, \pi \left[ \frac{77 096 675}{254 016} + \frac{378 515}{1 512}
    \eta - \frac{74 045}{756} \eta^2 \right] \nonumber \\  & &
  \left. (\pi M_z f)^{7/3}  \right] \,
  ,\label{eq:PNpsi} \end{eqnarray}
\noindent where $M_z = (1+z) (m_1 + m_2)$ is the binary's redshifted
total mass, $\eta = \mu_z / M_z$ is defined as the binary's symmetric
mass parameter, $\gamma$ is Euler's constant, and $f_0$ is a constant
frequency scale \citep{blanchet06}.
Central to the results of this
paper and N10, key geometrical source parameters, such as $D_L$ and $\cos\iota$, appear in the amplitude of
each GW polarization $\tilde{h}_\times(f)$ and
$\tilde{h}_+(f)$. Therefore, measurement errors in $D_L$ and $\cos\iota$ depend on the extent of the degeneracy
between these and other parameters appearing only in the amplitude. We hence
wish to assess how well we can disentangle each polarization from the measured GW strain at a
detector. 

Beyond the redshifting of masses (which is a simple consequence of the
cosmological redshift of all timescales), this waveform model does not
encode any information about source redshift. To investigate the $D_L$--$z$ relationship,
in this work we require an independent measure of the source's $z$ by observing an EM
counterpart. Other methods of obtaining the source's redshift
include using statistical arguments regarding the underlying NS binary merger
distribution (e.g.,~\citealt{Taylor:2012}), or adding information
about the NSs' (non-redshifted) tidal deformation in the GW phase
(e.g.,~\citealt{Messenger:2012}). \cite{DelPozzoPRD:2012} uses galaxy
catalogs to infer probabilistically sources' redshifts. In
contrast, an EM counterpart detected with a GW measurement may also
advantageously indicate the source's sky position. As was shown in N10, localizing the
binary with independent EM observations reduces 
measurement errors in parameters $D_L$ and $\cos\iota$ by breaking correlations with other parameters, and
by reducing the dimensionality of the parameter space. An EM counterpart may also
bring information about the time of merger for the
binary, which will increase the detection
range of a coherent network by a factor
of $\sim 1.2$ (e.g., \citealt{KelleyPRD:2013,Dietz:PRD2013}).

\section{Method}
 
This section summarizes the methodology used to derive $H_0$
measurements for an ensemble of NS-NS or NS-BH binary mergers. We first outline the schema of our method. Based on
Sections 3 and 4 of N10, we then describe technical aspects of simulating anticipated distance measurements.
 
\subsection{Schema of our method}

We detail below how we construct the posterior probability density function (PDF) in
$H_0$ for a set of detected GW-EM standard siren
measurements.
Deriving $H_0$ constraints for ensembles of NS binary
mergers requires particular care, as we expect that the majority of events
will be detected at low SNR. Consequently, PDFs in $H_0$ for
individual events will depend significantly on our prior knowledge of
the events' parameter distributions. In this study, how we select for
GW-EM events determines our choice in
specific priors.

We envision a scenario in which we have detected a total of $m$ GW-EM
events. Each event has both a GW measurement of distance, $D_{L,i}$, and
an EM redshift, $z_i$, where the subscript $i$ represents a particular binary
and runs from $1 \ldots m$. When combined these produce a value for
$H_0$ [see Eq. (\ref{eqn:DLz})]. We assume a model that is described by: i) the event's underlying
redshift distribution denoted by $X$, ii) each source's \emph{true}\/
redshift $\hat{z}_i$, and iii) the vector set of source
parameters $\boldsymbol{\theta^R}_i$ for a single measured GW binary.
As shown in Eq.~(\ref{eqn:DLz}), the luminosity distance
for a specific GW event depends on both $H_0$ and $\hat{z}$, and thus, we do \emph{not}
include $D_{L,i}$ in $\boldsymbol{\theta_i^R}$. The set
$\boldsymbol{\theta^R}$ differs from the set
$\boldsymbol{\theta}$, which includes the
parameter $D_L$ and was used in N10. In N10 we were interested in luminosity distance
measurements for individual events and
not for ensemble GW-EM standard sirens as in this work.

The data matrix $\{ \mathbf{s}_i, z_i \}$ comprises the measured GW
time streams $\mathbf{s}_i$, and the set of \emph{observed} EM
redshifts $z_i$ for a set of $m$ binaries.  If we assume that our model parameters are independent of one another, the
prior $p_{\mathrm{prior}} \, (H_0, X, \hat{z}_i,
\boldsymbol{\theta^R}_i )$ for $m$ detected GW-EM coincident
events is given by:
\begin{eqnarray} 
p_{\mathrm{prior}} \, (H_0, X, \hat{z}_i, \boldsymbol{\theta^R}_i) & = & {\cal A} \,
\, p_0 (H_0) \, \, p_0 (X) \nonumber \\ \times
& & \, \, \prod_{i=1}^{m} \, \, p_0 (\hat{z}_i
| X) \, \, p_0 (\boldsymbol{\theta^R}_i) \, \, ,
\label{eq:combinedprior}
\end{eqnarray}
where $p_0 (H_0)$, $p_0 (X)$, and $p_0 (\boldsymbol{\theta^R}_i)$  are
the individual priors on $H_0$, $X$, and
$\boldsymbol{\theta^R}_i$. The quantity $p_0 (\hat{z}_i
| X)$ is the prior distribution on a GW-EM event's true redshift
given the underlying distribution $X$, and ${\cal A}$ is a
normalization constant. We introduce the likelihood function ${\cal
  L}$, which measures the relative conditional
probability of observing the sources' redshifts $z_i$ (via EM measurements), and a particular set
of data $\boldsymbol{s}_i$ (via GWs) given the source's
parameters $\boldsymbol{\theta^R}_i$. It assumes the form:
\begin{eqnarray}
{\cal L} ( \{ \boldsymbol{s}_i, z_i \} | H_0, X, \hat{z}_i ,
\boldsymbol{\theta^R}_i ) & = & \prod_{i=1}^{m} {\cal L} (\mathbf{s}_i |
H_0 , z_i , \boldsymbol{\theta^R}_i) P(z_i | \hat{z}_i) \, \, ,
\nonumber \\ & & 
\label{eq:likelihood}
\end{eqnarray} 
\noindent
where the likelihood function for a single GW event is 
\begin{eqnarray}
{\cal L} (\mathbf{s}_i | H_0 , z_i , \boldsymbol{\theta^R}_i) & =  & \, e^{ -
\big( h_a({\boldsymbol \theta}) - s_a \, \big| \, h_a ({\boldsymbol
\theta}) - s_a \big)/2 } \, .
\label{eq:Likeexplicit}
\end{eqnarray}
The inner product $\left( g | h \right)$ describes the noise-weighted
cross correlation of $g(t)$ and $h(t)$ on the vector space
of signals, and is defined as:
\begin{eqnarray}
(g|h) & = & 2 \int_0^{\infty} df \frac{\tilde{g}^*(f)\tilde{h}(f) +
\tilde{g}(f)\tilde{h}^*(f)}{S_n(f)} \, ,
\label{eq:innerproduct}
\end{eqnarray}
where $S_n(f)$ denotes the instrument's power spectral
density. The Fourier transform ${\tilde h}(f)$ of $h(t)$ is defined as:
\begin{equation}
{\tilde h}(f) \equiv \int_{-\infty}^{\infty}\, e^{2\pi i f t}h(t)\, dt\;.
\label{eq:fourierT}
\end{equation}

An important element of our analysis is that we express the 
\emph{joint posterior}\/ PDF in $H_0$ given $m$ observed GW-EM events as:
\begin{eqnarray}
\nonumber
p_{\mathrm{joint}} \, (H_0 |\{ \mathbf{s}_i, z_i \})  & \propto & \,
\, 
p_{\mathrm{prior}} \, (H_0, X, \hat{z}_i, \boldsymbol{\theta^R}_i)
\nonumber \\ & & \times \, {\cal L} ( \{ \boldsymbol{s}_i, z_i \} | H_0, X, \hat{z}_i ,
\boldsymbol{\theta^R} ), \nonumber \\ \nonumber
& = & \, {\cal N} \, \, p_0 (H_0) \int dX \, p_0 (X) \\ & &
\bigg \{ \prod_{i=1}^{m} \int d\hat{z}_i \, \, p_0(\hat{z}_i | X) \,
P(z_i | \hat{z}_i)  \nonumber \\
 & &  \times \bigg[ \int d \boldsymbol{\theta^R}_i \, \,  p_0
   (\boldsymbol{\theta^R}_i)  \, \nonumber \\ 
& & \, \times  {\cal L} (  \mathbf{s}_i | H_0, X,
   z_i , \boldsymbol{\theta^R} )  \bigg] \bigg \} \, ,
\label{eq:jointPDFH0}
\end{eqnarray}
\noindent
where ${\cal N}$ is a normalization constant and we substitute
Eqs.~(\ref{eq:combinedprior}) and~(\ref{eq:likelihood}) for the
model's prior and likelihood functions respectively. In the event where we
have the precise redshift measurement of binary $i$'s EM counterpart
[i.e., $p(z_i | \hat{z}_i) = \delta(z_i - \hat{z}_i)$], Eq.~(\ref{eq:jointPDFH0})
then reduces to:
\begin{eqnarray}
\nonumber
p_{\mathrm{joint}}(H_0 |\{ \mathbf{s}_i, z_i \}) & = & {\cal N} \,
p_0 (H_0) \int dX \, p_0 (X) \bigg\{ \prod_{i=1}^{m} p_0(z_i | X) \nonumber \\
& & \times \bigg[ \int d \boldsymbol{\theta^R}_i  \, \, p_0
(\boldsymbol{\theta^R}_i) \nonumber \\  & & \left. \times  {\cal L} (
  \mathbf{s}_i \, | \,  H_0, X, z_i , \boldsymbol{\theta^R}_i ) \bigg]
  \bigg\}   \nonumber \right. \\ & = &  {\cal N} \, p_0 (H_0) \left[ \, \int dX p_0 (X)
  \prod_{i=1}^{m} p_0(z_i | X) \right] \nonumber \\ 
& & \times \prod_{i=1}^{m} \bigg \{ \int d \boldsymbol{\theta^R}_i \, p_0
(\boldsymbol{\theta^R}_i) \,  \nonumber \\ & & \times \,  {\cal L} (  \mathbf{s}_i | H_0, X, z_i , \boldsymbol{\theta^R}_i ) \bigg \}\nonumber \\ 
&= & \,  {\cal N}' \, p_0 (H_0)  \, \times \prod_{i=1}^{m} \bigg \{ \int d
\boldsymbol{\theta^R}_i  \, \, p_0
(\boldsymbol{\theta^R}_i) \, \nonumber \\ & & \times \, {\cal L} (  \mathbf{s}_i | H_0, X, z_i , \boldsymbol{\theta^R}_i
) \bigg \} \, .\label{eq:jointPDFH0precise}
\end{eqnarray}
The normalization constant ${\cal N}'$ absorbs the $[ \ldots ]$ part
appearing in the previous line, which is independent of $H_0$. We
assume a uniform prior in $H_0$ such that $p_0(H_0) = $ constant.  It is worth noting that our formalism could be generalized to include the most precise current estimates of $H_0$.  We do not do this here, although we certainly imagine that this would be done when one does an analysis of this sort with actual GW detections.

Since we take $p_0$ to be constant, we need only compute the $\{ \ldots \}$ term
in the last part of Eq.~(\ref{eq:jointPDFH0precise}), where $\boldsymbol{\theta^R}_i$ 
does not include $D_L$. Outlined below, our work relies on computing
the key term:
\begin{equation}
\label{exp:priorLike} 
 p_0
(\boldsymbol{\theta^R}_i) \, \, {\cal L} (  \mathbf{s}_i | H_0, X, z_i ,
\boldsymbol{\theta^R}_i ) \, \nonumber 
\end{equation}
for each GW-EM event. This contrasts with the methods
used in N10, where we instead computed the term $\, p_0
(\boldsymbol{\theta}_i) \, \, {\cal L} (  \mathbf{s}_i |
\boldsymbol{\theta}_i ) \, \, $ for each binary, where
$\boldsymbol{\theta}$ included $D_L$.

\subsection{Summary of MCMC approach used}

For each GW-EM event we explicitly map out the term $ p_0 (\boldsymbol{\theta^R}_i) \, {\cal L}_{\rm TOT} (  \mathbf{s}_i | X, z_i ,
\boldsymbol{\theta}_i )$ using MCMC methods (see N10 and
\citealt{Nissanke:2011} for details). The vector set $\boldsymbol{\theta}$
now includes $D_L$ because of the one-to-one mapping between $D_L$ and $H_0$. The quantity ${\cal L}_{\rm TOT} (  \mathbf{s}_i | X, z_i ,
\boldsymbol{\theta}_i )$ is the likelihood function for an entire
network. We assume that the instrument noise ${\bf n}$ is Gaussian, independent, and uncorrelated at each detector
site. Therefore, the network likelihood function is the product of
the individual likelihoods at each detector. We generate the signal
${\bf s}_a$ at each detector $a$ such that it
comprises the predicted GW signal ${\bf h}_a (\boldsymbol{\hat{\theta}})$,
which depends on the set of true source parameters
$\boldsymbol{\hat{\theta}}$, and the instrument noise ${\bf n}_a$.  In
our study we use the projected advanced LIGO sensitivity curve for $S_n(f)$ shown in N10 and
denoted ``Zero--Detuned, High--Power'' in \cite{ligo_noise} for
all our GW interferometers.

We generate predicted templates $\mathbf{h}_a$ (and hence also the
measured signals $\mathbf{s}_a$) using the PN description of the binaries as the bodies inspiral about one another prior to
their merger. Specifically, we use the restricted 3.5PN waveform in the frequency domain, where the GW frequency evolves with a characteristic
chirp [see Eqns. (4)--(6)]. We note that the largest contribution to the
signal accumulates from the
inspiral (and not the subsequent merger and ringdown parts of the waveform) for NS binaries in the frequency band of
ground based interferometers (\citealt{fh98}). When the sky position
$\mathbf{n}$ is assumed known from its EM counterpart
observation, the GW strain at each detector,
$\mathbf{h}_a (t; \boldsymbol{\hat{\theta}})$, is described using seven
parameters, $\boldsymbol{\theta}$: the two redshifted mass
parameters (${\cal M}_z$ and $\mu_z$), two orientation angles ($\Psi$
and $\cos \iota$), the GW merger's time and phase ($t_c$ and
$\Phi_c$), and the binary's luminosity distance ($D_L$). Apart
from excluding $D_L$, the six parameters in the reduced vector set $\boldsymbol{\theta}^R$ are identical to
those in $\boldsymbol{\theta}$.

We use the MCMC algorithm discussed in Section 3.3 of N10 to explore
the likelihood function. For binaries with an underlying population with isotropic orientation, we take prior
distributions in the sources' parameters to be flat over the region of
sample space that corresponds to our threshold SNR (described below). For the subset of beamed
binaries, we assume a uniform prior on the SGRB's beaming angle
distribution in the range of $| \cos \iota | > 0.94$, which corresponds
to a beamed population with an opening jet angle of approximately
$20^{\circ}$ (e.g., \citealt{Burrows:2006,Soderberg:2006,Fong:2012}). We choose the prior such that it
is fully consistent with how we select our subset of beamed SGRBs. 

\subsection{Binary Selection}

We follow the approach of N10 in generating a sample of detectable GW-EM
events. Having assumed a constant comoving density of GW-EM events in a $\Lambda$CDM universe \citep{komatsu09}, we distribute $10^6$
binaries uniformly in volume with random sky positions and
orientations to redshift $z = 1$ ($D_L \simeq 6.6$ Gpc). 

By computing the expected network SNR (the root-sum-square of the
expected SNRs at each detector) for each binary and comparing it to
a threshold network SNR, we construct a \emph{detected}\/ sample of
binary events for every network under consideration.\private{Using the expected SNR
instead of the observed SNR at a detector will introduce a small error for each
binary; the net effect for an ensemble of binaries should be
negligible.} Assuming prior knowledge of merger time and
source position allows us to set the threshold for the network to $\mbox{SNR} = 7.5$, a value
lower than that used in the absence of an EM counterpart (see D06 and N10). Figure 2 in N10
shows how the detectable GW-EM events for each detector network increases as the number of detectors in a network
increases. Notice that N10 included an advanced detector in Australia
and not LIGO India as in this work (the coordinates assumed for LIGO
India are given in \citealt{Nissanke:2012b}). The term ``total detectable binaries'' refers to binaries which
are detectable by a network of all five detectors --- both LIGO
sites, Virgo, LIGO India, and KAGRA.

Finally we obtain our subsample of beamed SGRBs from our original sample of
total detected GW-EM binaries by assuming that the SGRB has a uniform beaming
angle distribution of $|\cos \iota| > 0.94$.


\section{Results and Discussion}


We now present our results for $H_0$ constraints using
GW standard sirens observed by advanced ground-based GW detector networks. As discussed in
Sec.~2, we are interested in the joint PDF of $H_0$ given 
an ensemble of GW-EM observations. 

\begin{figure}
\centering 
\includegraphics[width=0.8\columnwidth]{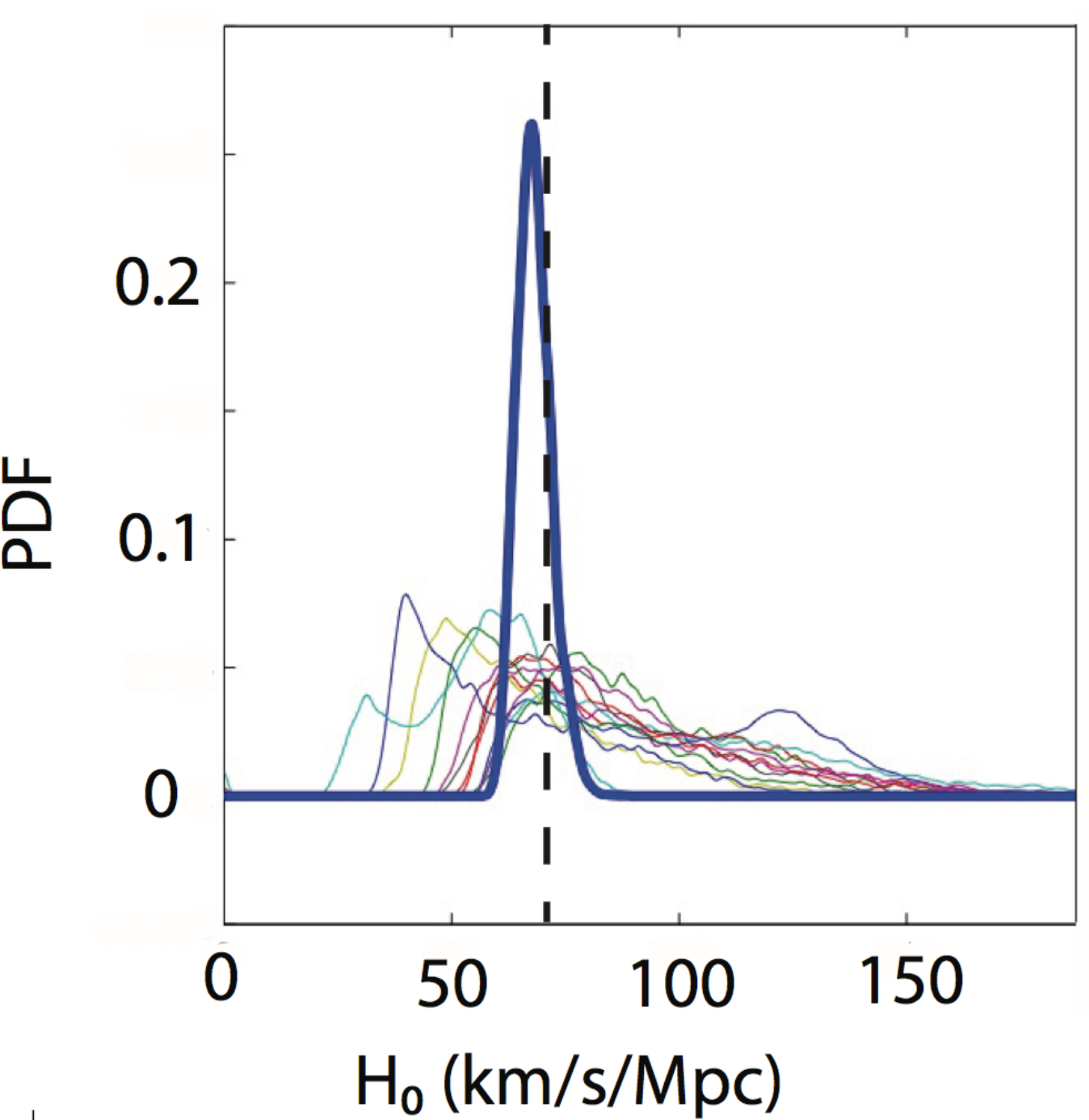}
\caption{The joint posterior PDF in $H_0$ for a sample of 15
  isotropically-oriented NS-NS binaries observed using a three
detector network (LIGO Livingston, LIGO Hanford, Virgo). The light coloured lines mark
the normalized posterior PDF for $H_0$ for each event, whereas the
thicker blue line denotes the joint posterior PDF in $H_0$ given all
the observed events. The vertical dashed black line denotes the value of $H_0$ of $70.5 \,
\mathrm{km}/\mathrm{s}/\mathrm{Mpc}$ used in generating the simulations. As the number of detections
increases, the joint posterior PDF gets progressively narrower, and its center comes closer to the true value of $H_0$.}
\label{fig:H0NSNSdet}
\end{figure}

Figure~\ref{fig:H0NSNSdet} shows the normalized joint posterior
PDFs in $H_0$, indicated by the thick blue line, for a sample of 15 isotropically oriented NS-NS
binaries observed with the baseline LIGO-Virgo network. The
thin lines in Figure~\ref{fig:H0NSNSdet} represent each
individual binary's measurement of $H_0$. We note that each binary
gives a relatively poor constraint on $H_0$ with 68\% confidence level (c.l.) fractional errors of
$\sim 30$--$50 \%$. The likelihood formalism results in the joint posterior
PDFs in $H_0$ for an ensemble of 15 mergers detected by LIGO-Virgo network being peaked around our assumed true
value of $70.5 \,  \mathrm{km}/\mbox{s}/\mathrm{Mpc}$ with a standard
deviation of $5 \,  \mathrm{km}/\mbox{s}/\mathrm{Mpc}$. This result would be
competitive with current $H_0$ constraints using either the
cosmological distance ladder (e.g., see \citealt{Freedman:2012},
\citealt{Riess:2009a}a, \citealt{Riess:2009b}b) or VLBI water maser
measurements (see e.g., \citealt{ReidApJ:2009,BraatzApJ:2010,KuoApJ:2011,ReidApJ:2013}). 

Table~\ref{tab:H0errors} presents the errors (68\% c.l.) on $H_0$ as a function of
the number of binaries detectable by a particular
network. We randomly select 30 NS-NS and 30 NS-BH mergers detected in
GWs using a five detector network. The actual number of detectable binaries is a
function of the detector network, on whether the EM counterpart is collimated (as we expect in the case
of SGRBs), and on whether the progenitor model is a NS-NS or NS-BH
binary~(N10, \citealt{Chen:2012}). Specifically, Table~\ref{tab:H0errors} shows the
measurement errors in $H_0$ for samples of unbeamed and beamed NS-NS
or NS-BH binaries using different detector networks. The percentage errors are
quoted as the fraction of measured standard deviations over an
assumed true value ($H_0 = 70.5 \,
\mathrm{km}/\mathrm{s}/\mathrm{Mpc}$). Such a measurement corresponds to a range of observation times because
of the wide range of uncertainties in NS binary merger rates. The general trends seen in
Table~\ref{tab:H0errors} can be summarized as:
\begin{itemize} 
\item As expected, the errors in $H_0$ decrease with
an increase in the number of GW detectors in a network. Table~\ref{tab:H0errors}
shows that a five detector network will result
in an improvement of up to a factor of 2 compared to a
three detector network.  Such a
feature is a consequence of the increase in the \emph{detected}\/
number of binaries, rather than due to the decrease in measurement error in $D_L$ for
each individual event (see discussion in N10). Due to differences in
the instruments' antenna response functions, the addition of LIGO India has a greater impact
than that of KAGRA.

\item The errors in $H_0$ reduce by a factor
  from two to five when the EM counterpart is assumed to be beamed. In the case of beamed NS binary
  mergers whose PDFs are more Gaussian in shape, we find that the error in $H_0$ decreases as $1/\sqrt{N}$, where
  $N$ is the number of GW-EM events detected. We expect this trend in the joint
  PDF of $H_0$ as we fix the
  inclination angle of each binary in the ensemble, since individual $H_0$ constraints are Gaussian in
  distribution due to the absence of the $D_L$-$\cos \iota$
  degeneracy.

\end{itemize}
  
\begin{deluxetable*}{lccccc}
\tabletypesize{\scriptsize}
\tablewidth{16.0cm}
\tablecaption{Measurement errors in $H_0$ for a sample of
GW-EM events. Results are presented for unbeamed and beamed sources, for both
NS-NS and NS-BH mergers, and for a range of
detector networks. The $\%$ values are the 68\% c.l. fractional errors, and the number of binaries detected by each 
network is given in parentheses.}
\tablehead{
\colhead{Network} &
\colhead{LIGO+Virgo (LLV)} &
\colhead{LLV+LIGO India} &
\colhead{LLV+KAGRA} &
\colhead{LLV+LIGO India+KAGRA} &
}

\startdata

NS-NS Isotropic & 5.0\% (15) & 3.3\% (20)  & 3.2\% (20) & 2.1\% (30)  \\
& & & & \\
NS-NS Beamed & 1.1\% (19)  & 1.0\% (26) & 1.0\% (25) & 0.9\% (30) \\
& & & & \\
NS-BH Isotropic & 4.9\% (16) &  3.5\% (21) & 3.6\% (19) & 2.0\% (30) \\
& & & & \\
NS-BH Beamed & 1.2\% (18)  & 1.0\% (25) & 1.1\% (24) & 0.9\% (30) \\
\enddata

\label{tab:H0errors}
\end{deluxetable*}

\begin{figure}
\centering 
\includegraphics[width=0.98\columnwidth]{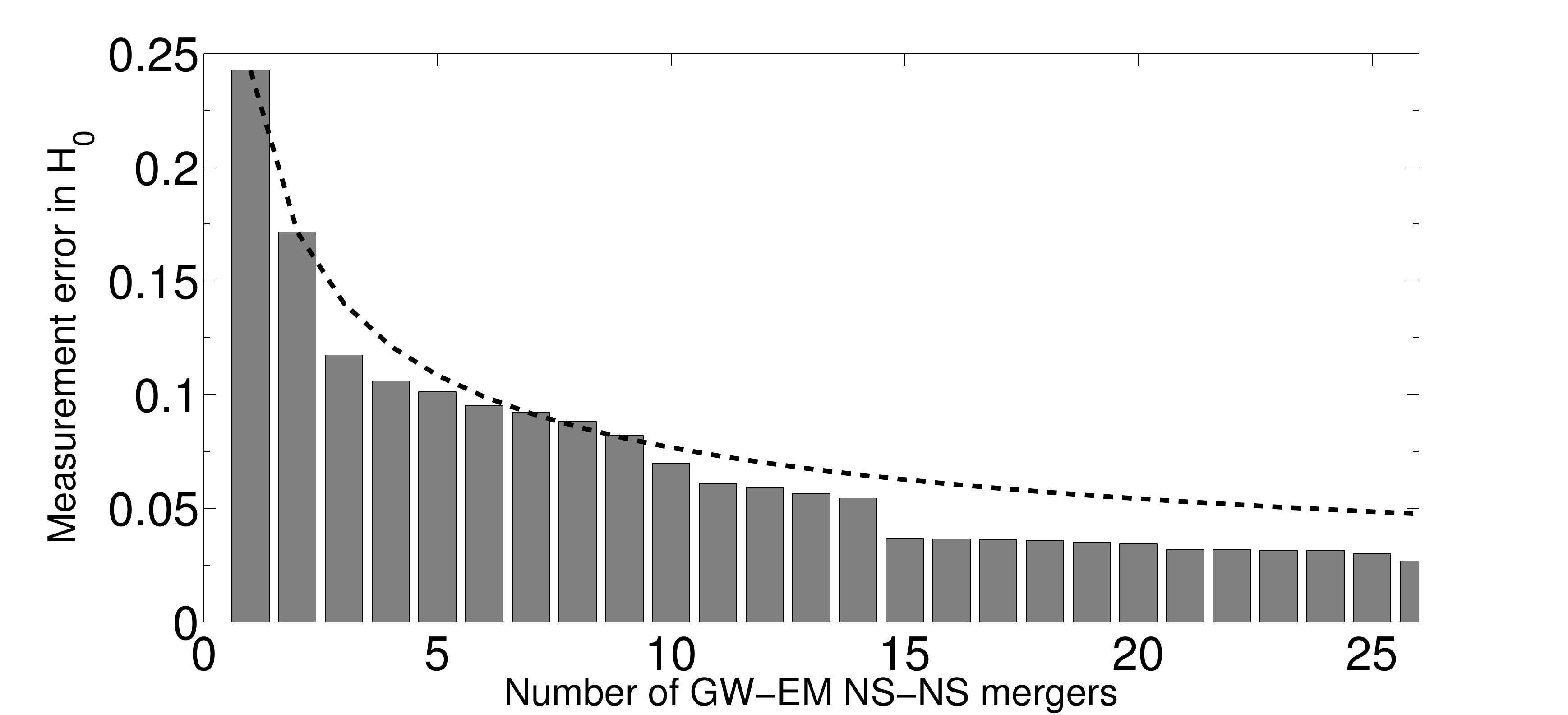}
\caption{$H_0$ measurement error as a function of the number of multi-messenger
  (GW+EM) NS-NS merger events observed by a LIGO-Virgo network. The
solid bars indicate the 68\% c.l. measurement error in $H_0$ for
the {\it joint} PDF of the independent binary mergers; the dashed line shows the 68\% c.l. measurement error in
$H_0$ derived assuming Gaussian errors for each GW-EM merger. When
specifying the particular order of events shown, we
choose the GW-EM merger in the remaining ensemble with the mean measurement error in $H_0$.}
\label{fig:H0fntime}
\end{figure}

Fig.~\ref{fig:H0fntime} shows the 68\% c.l. measurement error in $H_0$ as a function of
the number of GW-EM detectable NS binary merger events. We assume that
our detectable sample comprises 26 GW-EM binary
mergers observed with a LIGO-Virgo network; we expect that the
errors in $H_0$ will decrease with $1/\sqrt{N}$ in the limit of large
$N$, where $N$ is the number of detectable GW-EM events. We compute the
posterior PDF in $H_0$ for each NS-NS
binary merger in our sample averaged over 100 noise realizations. The
solid bars indicate the measurement error in $H_0$ for the {\it joint} PDF of some $i$ binary mergers; at
low $i$, we select the $i$-th merger with the mean value in the $H_0$
error of the remaining $(26-(i-1))$ detectable GW-EM events. By doing so, we minimize the
impact that arbitrary ordering for small $i$ GW-EM events will have on the convergence
of measurement errors in $H_0$. For an identically-ordered ensemble of
NS-NS mergers, the dashed line indicates the measurement error in
$H_0$ derived assuming Gaussian errors for each GW-EM independent
merger. In the limit of large $i$ events, the difference in $H_0$
error constraints decreases between the two methods. Furthermore, for low $i$
events in particular, we find that the non-Gaussian
shapes of the individual $H_0$ distributions improve the combined
$H_0$ distribution. For example and as discussed in N10, after observing 15 NS-NS mergers in
GWs and EM, we find that $H_0$ may be measured to within 5\% using the combined posterior PDF method (or
to within 8\% assuming Gaussian posterior PDFs for each individual
event). Without an EM counterpart and based solely on statistical
cross-correlations of GW sky errors with wide-field galaxy surveys,
\cite{DelPozzoPRD:2012} finds a 14\% $H_0$ measurement error (with a 95\% confidence interval)
using ten GW merger events with a LIGO-Virgo network (and assuming a
SNR $\sim15$). 

We now explore how the number of detectable GW-EM NS binary merger events corresponds to an
observable time window. From \cite{Abadieetal:2010}, we use the
mean NS-NS merger rate of 1 Mpc$^{-3}$ Myr$^{-1}$. We expect 15 (30)
isotropically-oriented NS-NS mergers to be detectable in GWs over
a $\sim$ three month period using a three (five) GW detector network and an EM precursor
trigger \citep{Nissanke:2012b}.  If we instead consider beamed NS binary mergers and use the
SGRB rate of 10 Gpc$^{-3}$ yr$^{-1}$, we expect $\sim$ 30 GW-SGRB
events per year \citep{Berger:2011,Chen:2012,EnricoPetrilloApJ:2013}. 

In the case of isotropically-oriented NS-10 M$_{\odot}$ BH mergers, we
use a merger rate of 0.03 Mpc$^{-3}$ Myr$^{-1}$
\citep{Abadieetal:2010}. We then expect 15 (30) GW-detectable events in GWs over a six month period
(we scale the results given in Table 1 of \citealt{Nissanke:2012b} by
a factor ${\cal M}_c^{5/6}$ to account for the difference between the
NS-10 M$_{\odot}$ BH and the NS-5 M$_{\odot}$ BH mergers used there
and here respectively). Due to an
absence of observed systems, we emphasize that NS-BH merger rates
based on population sythesis results vary by several orders of
magnitude. In the case of beamed NS-BH mergers, we use
the SGRB rate of 10 Gpc$^{-3}$ yr$^{-1}$ and find 1 GW-SGRB
event per year (e.g., \citealt{Chen:2012,EnricoPetrilloApJ:2013,KelleyPRD:2013,Dietz:PRD2013}).

Although it is clear that of order 20--30 events are needed to reach
percent level accuracy in determining $H_0$, it is unclear how long this will
take given the range of uncertainty in binary merger rates. Current estimates suggest that the median timescale to
achieve this number of events is likely about one year, but could be
as short as a few months, or as long as a decade.

\section{Implications for Cosmology}

Assuming GR accurately describes the inspiral dynamics and GW
emission, GW standard sirens should provide a measure of
$H_0$ based on \emph{absolutely-calibrated}\/ GW distances that are independent of the
cosmological distance ladder. Given that we anticipate a network of
advanced GW
interferometers reaching their design sensitivity within the next
decade, this physics-based technique could play a large role in
precision determination of the Hubble constant, especially in conjuction with
other approaches~(see
\citealt{Suyu:2012} and references therein).

In this work, by envisioning a range of scenarios using different networks of GW
detectors and different populations of NS binary progenitors, we show that
ensembles of GW standard sirens have the power to constrain $H_0$ to an accuracy of $\sim 1
\%$. We have assumed joint GW and EM observations of the NS binary
merger; other works, for instance \cite{Taylor:2012},
\cite{DelPozzoPRD:2012} and \cite{Messenger:2012}, examine $H_0$ constraints using solely GW
observations, and are based on statistical arguments or galaxy catalogs to
infer the mergers' redshifts. We emphasize that an individual standard siren may only constrain $H_0$
to a precision ranging from 5 to 50\%.
We have shown that the error in $H_0$ depends critically on \emph{the number of GW-EM mergers observed}, which in turn
depends on the NS binary progenitor, on whether the NS binary is face-on (due to
GRB beaming), and on the number and sensitivy of GW interferometers in a network. We find that the
critical limitation when projecting the timescale for this measurement (once the
GW detectors are operational) is the few orders of magnitude
uncertainty in NS binary merger rates, independent of GW detections. Using mean
NS merger rates derived from population synthesis or the observed Galactic
binary pulsar distribution, we estimate that percent-level measurements of $H_0$
are possible within $\sim1$ year of observation, or may take as
long as a decade for pessimistic event rates.

For flat cosmologies, a measurement of $H_0$ at the percent level, when combined
with precision CMB measurements of 
the absolute distance to the last scattering surface, would constrain the dark energy equation of
state parameter $w$ to $\sim 10\%$ (D06). The power of such a result (e.g., to
falsify the cosmological constant model for dark energy) depends critically on
understanding the systematic errors associated with the measurement of
$H_0$. It is for this reason that GW standard sirens may have an
important role to play in constraining cosmology in the near future.

\section{Acknowledgements}


We thank Curt Cutler, Phil Marshall, and Michele
Vallisneri for very useful discussions on selection
effects and biases. We thank Vicky Scowcroft for discussion on $H_0$
measurements, Edo Berger, Josh Bloom and Brian Metzger
for discussions on GW-SGRB measurements, and Francois Foucart for discussion on the status of
numerical relativity simulations. Some of the simulations were performed using the
Sunnyvale cluster at Canadian Institute for Theoretical Astrophysics (CITA), which is funded by NSERC and CIAR. Part of this work was performed at the
Jet Propulsion Laboratory, California Institute of Technology, under contract
with the National Aeronautics and Space Administration. SMN is
supported by the David \& Lucile Packard Foundation. ND is supported by NASA under grants NNX12AD02G and NNX12AC99G, and by a Sloan Research Fellowship from the Alfred P. Sloan Foundation. DEH acknowledges support from National Science
Foundation CAREER grant PHY-1151836. SAH is supported by NSF Grant
PHY-1068720. SAH also gratefully acknowledges fellowship support by
the John Simon Guggenheim Memorial Foundation, and sabbatical support
from CITA and the Perimeter Institute for Theoretical Physics. CH is supported by the Simons Foundation, the David \& Lucile Packard Foundation, and the US Department of Energy (award DE-SC0006624).

\newpage

\bibliography{sirens2}
\end{document}